\documentclass[twocolumn,showpacs,preprintnumbers,amsmath,amssymb,groupedaddress,superscriptaddress]{revtex4-1}

\usepackage{graphicx}
\usepackage{dcolumn}
\usepackage{bm}

\begin{document}

\title{Measurements of vector magnetic field using multiple electromagnetically induced transparency resonances in Rb vapor }
\author{Kevin Cox}
\affiliation{Department of Physics, College of William\&Mary, Williamsburg,
Virginia 23185, USA}
\author {Valery I.~Yudin} 
\affiliation{Institute of Laser Physics SB RAS, Novosibirsk 630090, Russia} \affiliation{Novosibirsk State
Technical University, Novosibirsk 630092, Russia}\affiliation{Novosibirsk State University, Novosibirsk 630090,
Russia}
\author {Alexey V.~Taichenachev}
\affiliation{Institute of Laser Physics SB RAS, Novosibirsk
630090, Russia} \affiliation{Novosibirsk State Technical University, Novosibirsk 630092, Russia}\affiliation{Novosibirsk State University, Novosibirsk 630090,
Russia}
\author{Irina Novikova}
\affiliation{Department of Physics, College of William\&Mary, Williamsburg,
Virginia 23185, USA}
\author{Eugeniy E. Mikhailov} 
\affiliation{Department of Physics, College of William\&Mary, Williamsburg,
Virginia 23185, USA}

\date{\today}

\begin{abstract}

We study dependence of electromagnetically
induced transparency (EIT) resonance amplitudes on the external magnetic field direction in a \emph{lin}$||$\emph{lin} configuration in ${}^{87}$Rb vapor.
We demonstrate that all seven resolvable EIT resonances exhibit maxima or minima at certain orientations of the laser polarization relative to the wave vector and the magnetic field. This effect can be used for development of a high-precision vector EIT magnetometer.

\end{abstract}


%


%

\maketitle

The ability  to   measure  magnetic   field  with   high  precision   and  good
spatial  resolution  benefits  many  applications.  For  example,  detection
of   weak   magnetic a  field   distribution  gives   a   new   non-invasive
diagnostic  method  for  heart  and  brain  activities,  allows identification  of
defects  in  magnetizable  coatings  and   films,  and  can possibly be  used
for  a non-demolishing  readout  of stored  memory  domains. Many  magnetic
sensors  available   today,  such   as  SQUIDs   (superconducting  quantum
interference  devices)~\cite{squid},  spin-exchange relaxation-free  (SERF)
magnetometers~\cite{romalisNature03,Ref:BudkerRomalisReview},   and    various
optical pumping  magnetometers~\cite{alexandrovOE92} are sensitive  only to
the  magnitude of  the  magnetic field.  These magnetometers lose valuable information about magnetic field direction and  can allow reduced accuracy due to  ``heading
error'' in some systems~\cite{magnsensors}.

\begin{figure}[h]
\includegraphics[width=0.7\columnwidth]{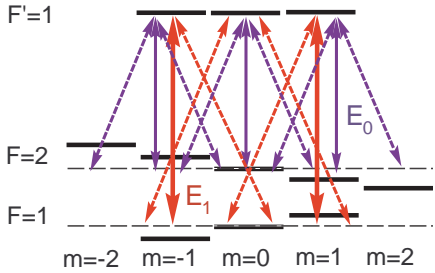}
\caption{
	Possible EIT $\Lambda$ systems for different magnetic field
	orientations: solid arrows show $\Delta m = 0$ transitions, and dashed
	arrows show $\Delta m = \pm 1$ transitions.
	\label{fig:multiEIT}
}
\end{figure}

The application of coherent optical effects, such as electromagnetically     induced     transparency     (EIT)~\cite{marangosEITreview} for magnetic field detection offers exciting perspective for development of high-precision miniature magnetometers~\cite{fleischhauerPRA94,wynands02APB,xiaAPL06,NISTmagnetometer,  BudkerPinesPNAS08, bison09}.  EIT  resonances are  associated
with the preparation  of atoms  into  a  coherent non-interacting  (dark)  superposition of  two metastable  states  of  an atom  (such as  two  Zeeman or  hyperfine  sublevels  of  the electronic  ground  state
of  an  alkali  metal  atom)
under  the  combined  interaction  of  two  optical  fields  $\mathrm{E_{0,1}}$ in  two-photon
Raman resonance  in a $\Lambda$ configuration.  The resulting ultra-narrow (down to a  few  tens  of   Hz~\cite{budker98prl,erhard01pra}) transmission peaks  can  be  used  to
measure, \emph{e.g.},  a frequency difference between  two hyperfine energy
levels  in  Rb  or  Cs  without  use  of  microwave
interrogation, making  this  method  particularly   attractive  for
development  of miniature  atomic clocks~\cite{vanier05apb,knappe05oe}  and
magnetometers~\cite{NISTmagnetometer}. In the latter case, the magnitude of
a  magnetic field  can be  deduced  from the  spectral position  of an  EIT
resonance  in  a $\Lambda$  link  formed  between magnetic  field-sensitive
Zeeman sublevels. However, this method is not sensitive to the direction of
the magnetic field vector $\vec{B}$.

Several  publications  suggested  that   the  information  about  $\vec{B}$
direction   can  be   extracted  by   analyzing  relative   intensities  of
various  EIT peaks~\cite{wynandsPRA98,leePRA98}.  Yudin  \emph{et
al.}~\cite{yudinPRA2010}   showed that a recently studied {\em lin}$||${\em  lin} configuration
is  a  promising  candidate  for  EIT
atomic   clock   applications~\cite{linlin05,matisovPRA09,zibrovPRA10,mikhailovJOSAB10}. The   amplitude   of   the
magneto-insensitive EIT  resonance is sensitive to the magnetic field direction and has  a  universal  maximum when  the laser polarization vector,  $\vec{E}$, is
orthogonal to the  plane formed by the magnetic field  vector, $\vec{B}$, and
laser wave vector, $\vec{k}$. Thus, measureing the resonance amplitude while rotating the laser polarization should provide information about magnetic field direction. Moreover, since this effect is based only  on fundamental
symmetries of the problem, corresponding measurement  procedure does  not require  any assumptions  regarding
the  details  of the experimental arrangements (such as laser power and detuning).

In this Brief  Report we explore the possibility  to simultaneously measure
magnetic  field  magnitude and  direction  by  recording both the spectral positions  and
relative amplitudes of multiple  Zeeman-shifted EIT resonances that occur for a bicromatic linearly polarized laser field interacting  with ${}^{87}$Rb  atomic vapor  placed in  an external
uniform magnetic  field. In  this case EIT  resonance conditions are fulfilled in various possible $\Lambda$-systems formed between Zeeman sublevels of two hyperfine states of Rb atoms, shown  in  Fig.~\ref{fig:multiEIT}, for several two-photon detunings  $\Delta_\mathrm{HFS}+   n\mu_B  g  B$, where
$n=0,\pm1,\pm2,\pm3$, $\Delta_{\mathrm{HFS}}$  is the hyperfine splitting,
$g$ is the gyromagnetic ratio, $\mu_B$ is the Bohr magneton.  Other  parameters of  each
EIT  peak (amplitude,  width)  strongly depend  on the  mutual orientation  of
magnetic field,  light polarization  and wavevector directions  that define
Rabi  frequencies  of optical  fields  for  each $\Lambda$  link  according
to  the  selection  rules.  However,  a  universal  intrinsic  symmetry  of
this  problem~\cite{yudinPRA2010}  predicts  that all  EIT  resonance
amplitudes  must  exhibit  local  maxima or  minima  for  two  orientations
of  light  polarization $\vec{E}$:  when  $\vec{E}$  is orthogonal  to  the
$\vec{B}$-$\vec{k}$  plane, and  when it  lays within  that plane.  Here we
confirm  (both experimentally  and by  using exact  numerical calculations)
that such  universal extrema  exist for all  observed EIT  resonances, even
though  their  characteristic strength  (maximum  vs  minimum) will  change
depending  on direction  of the  magnetic  field. Since  the exact  angular
position of such extrema~\cite{yudinPRA2010}  does not depend on parameters
of  the  laser (such  as  laser  intensity),  they should  provide  accurate
information about magnetic field orientation.

\begin{figure}[htp]
  \includegraphics[width=1.0\columnwidth]{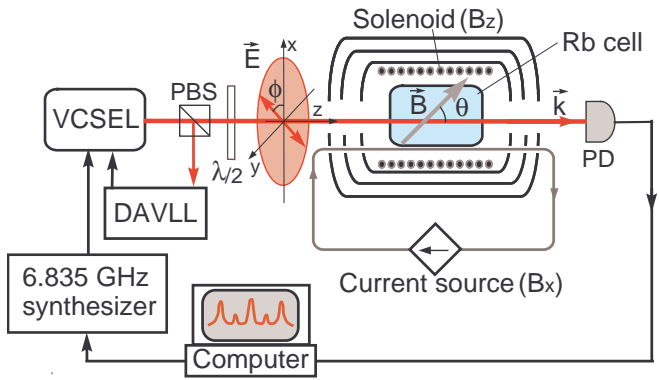}
  \caption{
  	\emph{(Color online)}
	Schematic of the experimental setup. See text for abbreviations.
	\label{fig:setup}
	}
	
\end{figure}

Our  experimental  arrangements   (see  Fig.~\ref{fig:setup})  are  similar
to   those   used   for    miniature   EIT-based   magnetometers~\cite{NISTmagnetometer}   (although
most  components  are   not  miniaturized).   The  details   of  the
construction  and   operation  of our  experimental  apparatus   are  given  in
\cite{mikhailovJOSAB10,belcherAJP09}.  A  temperature-stabilized
vertical  cavity   surface-emitting  diode  laser  (VCSEL) was  current-modulated  at
$\nu_m = 6.8347$~GHz  with the laser carrier frequency tuned  at  $5S_{1/2}F=2  \rightarrow 5P_{1/2}F'=1$  transition  and  the
first  modulation  sideband  resonant  with  $5S_{1/2}F=1  \rightarrow
5P_{1/2}F'=1$ transition (see  Fig.~\ref{fig:multiEIT}) using  a   dichroic-atomic-vapor  laser
lock  (DAVLL)~\cite{yashchukRSI00}. The intensity ratio
between  the  sideband  and  the  carrier  was  adjusted  by  changing  the
modulation  power  sent   to  the  VCSEL,  while   the  modulation  frequency
(and  consequently  the   two-photon  detuning  of  two   EIT  fields)  was
controlled by  a home-made  computer-controlled microwave  source operating
at  $6.835$~GHz~\cite{mikhailovJOSAB10}.  During  this experiment  we  kept the
sideband to  carrier ratio equal to  60\%, since this allows us to cancel the first
order  power shift  in this  setup~\cite{mikhailovJOSAB10,zibrovPRA10}. The
laser beam  with maximum total  power 120~$\mu$W and a  slightly elliptical
Gaussian  profile  [1.8~mm and  1.4~mm  full  width half  maximum  (FWHM)],
traverses a cylindrical Pyrex cell (length 75~mm; diameter 22~mm) containing
isotopically enriched $^{87}$Rb vapor and 15~Torr of Ne buffer gas, mounted
inside a three-layer magnetic shielding and actively temperature-stabilized
at $47.3^\circ$C. To control the polarization  of the laser before the cell,
the  beam passes  a polarizing  beam splitter  (PBS) and  then a  half-wave
plate ($\lambda/2$)  that rotates  the direction  of polarization  in $x-y$
plane. The  polarization angle  $\phi$ is  defined as  an angle  between the
polarization direction and a vertical $x$ axis.

\begin{figure}[h]
\includegraphics[width=0.9\columnwidth]{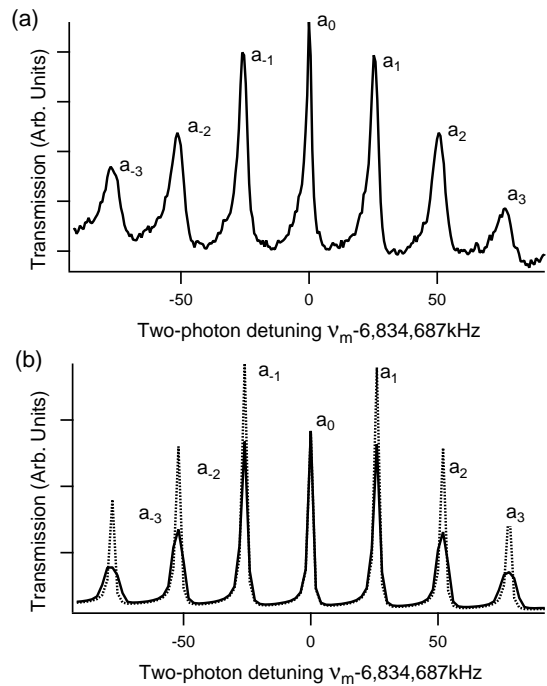}
\caption{
	\emph{(a)}Sample experimental EIT resonances for $\theta=90^\circ$ and $\phi=60^\circ$,
with angles defined in Fig.~\ref{fig:setup}. \emph{(b)} Matching theoretical calculations for a perfectly
    homogeneous magnetic field (dotted line) and taking into account transverse gradient from the wire (solid line).
     \label{fig:eit_sample}
}
\end{figure}

A  solenoid  inside  magnetic  shield  produces magnetic field $B_z$ parallel  to  $\vec{k}$, while a straight wire parallel to the
light propagation direction produces  a transverse component $B_x$. Special
care  was taken  to align  the laser  beam strictly  parallel to  the $B_x$
generating  wire  to avoid  variation  of  the  $B$  field along  the  beam
propagation  direction. Calibrated,  simultaneous adjustment  of current  in
both the solenoid  and the wire allowed  us to change the  direction of the
magnetic field $\theta$  (measured from the $z$-axis)  without changing the
magnitude of $B$ and associated 26~kHz Zeeman splitting.

During  experiments, for  a given  angles $\theta$  and $\phi$  we recorded
laser  field transmission  as a  function of  two-photon detuning by scanning the
laser's microwave modulation  frequency $\nu_m$ around the ${}^{87}$Rb hyperfine splitting.  A sample spectrum Fig.~\ref{fig:eit_sample}(a) shows seven  EIT
peaks that are labeled according to their Zeeman shift: the magneto-insensitive peak at $\nu_m=\Delta_{\mathrm{HFS}}$ is  labeled $a_0$, while
peaks separated by $\pm$  one, two and three  Zeeman splitting are
$a_{\pm1}$, $a_{\pm2}$,  and  $a_{\pm3}$ respectively.  Amplitudes  of  each EIT  peak  were
extracted  from the  Lorentzian  fit  and then  normalized  to the maximum
$a_0$ amplitude for each angle  $\theta$.

Our experimental results are supported  by the numerical calculations for complete hyperfine and Zeeman structure of Rb atoms based   on  the   standard  density-matrix   approach   (see,   \emph{e.g.},
Ref.~\cite{taichenachev03pra})  under  assumptions  of low  saturation  and
total collisional depolarization of the  excited state. The parameters used
in  the  calculations matched  the  experimental conditions:  total
intensity of the two fields $I_1$+$I_0$=0.5~$\mathrm{mW/cm}^2$ and $I_1/I_0=0.6$,
optical transition linewidth $\gamma=100$~MHz, ground   state   decoherence rate
$\gamma_0=500$~Hz, and Zeeman splitting $\mu_B g B=26$~kHz.

Fig.~\ref{fig:eit_sample}(b) shows a calculated EIT spectrum. To achieve good agreement with the experiment we took into account a transverse magnetic field gradient created by the straight wire that was characterized experimentally in our previous work~\cite{mikhailovOL}. Such variation of the magnetic field produced small differential shifts of the EIT resonance positions across the laser beam cross-section, that resulted in broadening of the magneto-sensitive peaks and change in their relative amplitudes. This effect was taken into account in numerical calculations by averaging the results over 25-point rectangular grid across the beam.

\begin{figure*}[h]
       \includegraphics[width=1.3\columnwidth]{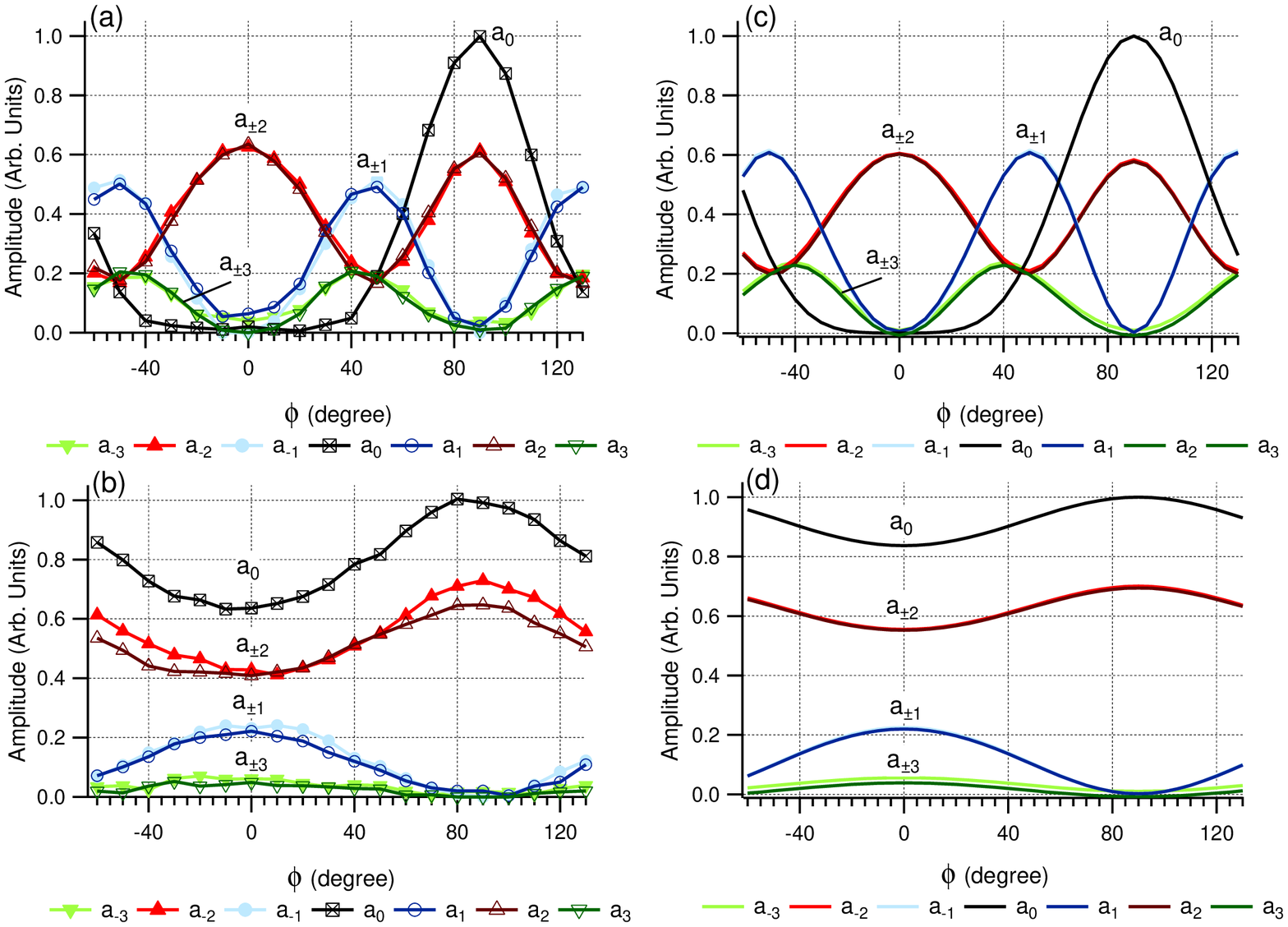}
       \caption{
       Experimental (a,b) and theoretical (c,d) dependence of the EIT
       resonance amplitudes on polarization angle $\phi$ for magnetic field
       angles $\theta=90^\circ$(top row) and $\theta=15^\circ$(bottom row).
	\label{fig:amplitudes_samples}
       }
\end{figure*}

Fig.~\ref{fig:amplitudes_samples} shows both experimental data and theoretical calculations for the amplitudes of all seven EIT resonances   as  functions  of  laser  polarization
angle   $\phi$  for two different
angles $\theta$.
The strongest dependence  on light polarization were observed for  mostly transverse magnetic
fields  $\vec{B}  \perp  \vec{k}$ [see Figs.~\ref{fig:amplitudes_samples}(a,c)].
According  to the selection  rules,   only  optical transitions  with $\Delta  m_F=0$ are
possible for $\vec{B}  || \vec{E}$  ($\phi=0^\circ$
and  $\theta=90^\circ$).  Thus,  only $a_{-2}$
and  $a_{2}$ EIT  resonances  appear  at  the two-photon  detunings
$\Delta_{\mathrm{HFS}}   \pm  2  g\mu  B$   (transition $F=1,m_F=0\rightarrow
F'=1,  m_F=0$ is  forbidden  due to  symmetry);  all other  EIT
peaks  vanish.
Similarly,  at  $(\vec{B}  \perp  \vec{k})  \perp  \vec{E}$
($\phi=90^\circ$  and  $\theta=90^\circ$),  only  transitions  with  $\Delta
m_F=\pm1$  are  allowed,   resulting  in  only  three   EIT  resonances  at
$\Delta_{\mathrm{HFS}}$  and  $\Delta_{\mathrm{HFS}}  \pm  2  g\mu  B$.  At
any  other intermediate  angle, the same  two Zeeman sublevels  can participate in more than one
$\Lambda$  systems, resulting  in constructive  or detractive  interference
depending  on  their Clebsch-Gordan  coefficients.  This  makes  calculations  of
EIT  resonance amplitudes  more complicated.  Furthermore, when there  is a  significant
longitudinal component of  the magnetic field present, the  amplitude of EIT
resonance  become less  sensitive to  the light  polarization, as  shown in
Figs.~\ref{fig:amplitudes_samples}(c,d).  However,  there are  still  clear
extrema at  $\phi=0$ and  $\phi=90^\circ$ for  all EIT  resonances. We experimentally observed this to be true for any direction of the magnetic field. The only exception is longitudinal magnetic field ($\vec{B}||\vec{k}$) since, as expected, no polarization dependence is observed.

In the ideally symmetric situation, the amplitudes of resonances with the same value but opposite detunings (\emph{e. g.}, $a_1$ and $a_{-1}$) should be identical. However, complex hyperfine and Zeeman structure of the excited state causes a slight asymmetry of EIT resonances (see Fig.~\ref{fig:eit_sample}) for nominal zero optical (one-photon) detuning. Combined  with  a relatively  small (compared  to  the resonance  linewidth) Zeeman  shift this asymmetry leads  to  a small
difference  in  resonance amplitudes for ``positive''  and
``negative'' peaks, which is more noticeable in our experiment for small angles $\theta$ [see
Fig.~\ref{fig:amplitudes_samples}(b)]. The resonance asymmetry can be reduced by optimizing the laser detuning~\cite{zibrovPRA10} and by operating at higher magnetic fields.

Measuring extrema positions of resonance amplitudes vs light polarization rotation angle dependence allows one to find the plane in which the magnetic field vector lies. For a complete measurement of magnetic field direction, one must take two independent measurements of two such planes, and then determine the direction of $\vec{B}$ from their intersect. This can be done, for example, by repeating the measurement for two
different light propagation directions in  the $x-z$ plane~\cite{yudinPRA2010}. Alternatively, an additional magnetic field of known magnitude and direction controllably ``rotate'' the total magnetic field for the second measurement.
However, for well-characterized experimental parameters the dependence of resonance  amplitudes  on polarization angle $\phi$  are unique  for every magnetic angle  $\theta$, forming a magnetic direction  ``fingerprint''. Such a fingerprint may allow a simplified procedure for extracting  $\vec{B}$ information.
Because  of  the  symmetry  of  the  {\em  lin}$||${\em  lin}  polarization
configuration,  the same  fingerprint  will be  obtained  for two  possible
direction combinations $\theta$ and $\pi-\theta$ with respect to  the $z$ axis.
This degeneracy  can  be lifted  by sending
an elliptically polarized light, which  brakes this symmetry
and makes  either positive  or negative resonances stronger depending on the magnetic field direction.


This measurement procedure can also be used to produce high-resolution maps
of  vector  magnetic fields  when used in  combination  with a  recently-demonstrated
magnetic field imager~\cite{mikhailovOL}, as long as light transmission across the laser beam is spatially
resolved. Since the  EIT resonance positions depends on  the Zeeman shifts
of  atomic  sublevels,  the  resonances will  occur  at  different  two-photon detunings  for different spacial locations in the case of a spacially varying  magnetic field.  Recording a series  of such  images for
various two-photon frequencies for a  particular EIT resonance can create a
spatial map of the magnetic  field magnitude. Repeating such measurements for several orientations of
the laser polarization and finding an extremas in resonance amplitude provides additional information about variations in the direction of the magnetic field.


In summary, we systematically studied 
dependence of multiple EIT resonances'
 on relative orientation of  magnetic field and laser polarization
in the {\em lin}$||${\em lin} configuration using a current-modulated VCSEL
on the  $D_1$ line of  ${}^{87}$Rb. We  demonstrated that all observed EIT
resonances can be used to extract complete information about magnetic field
vector with improved  sensitivity to directionality. These  findings can be
used  to  implement a  sensitive  small  scale vector  magnetometer  and/or
magnetic field imager with good spatial resolution.


The  authors  would   like  to  thank  S. Zibrov, A. S. Zibrov and V. L. Velichansky   for  useful  discussions.
This  research  was supported  by  the  National Science  Foundation  grant
PHY-0758010. A.V.T and V.I.Yu. were supported by RFBR (grants 09-02-01151, 10-02-00406, 11-02-00775, 11-02-00786, 11-02-01240),  DFG/RFBR (grant 10-02-91335), Russian Academy of Science, Presidium SB RAS, and by federal program ``Scientific and pedagogic personnel of innovative Russia 2009-2013''. K.C. acknowledges support from  the Virginia Space Grant Consortium through Undergraduate Research Scholarship.

\end{document}